\documentclass[]{emulateapj}
\usepackage{apjfonts}

\usepackage{epsfig, natbib, graphicx, color, bm}

\input epsf

\newcommand{\Mpc}{\mbox{Mpc}}
\newcommand{\hMpc}{h^{-1}\mbox{Mpc}}
\newcommand{\msun}{M_\odot}
\newcommand{\bx}{\bm{x}}

\newcommand{\nn}{\nonumber}

\newcommand{\avg}[1]{\left\langle #1 \right\rangle}
\newcommand{\Var}{\mbox{Var}}

\newcommand{\Hinv}{H^{-1}}

\newcommand{\Cov}{\mbox{Cov}}

\newcommand{\beq}{\begin{equation}}
\newcommand{\eeq}{\end{equation}}
\newcommand{\beqa}{\begin{eqnarray}}
\newcommand{\eeqa}{\end{eqnarray}}

\newcommand{\be}{\begin{equation}}
\newcommand{\ee}{\end{equation}}

\newcommand{\bea}{\begin{eqnarray}}
\newcommand{\eea}{\end{eqnarray}}

\newcommand{\Sc}{\Sigma_c}
\newcommand{\DS}{\widehat{\Delta \Sigma}}

\newcommand{\kpc}{\mbox{kpc}}

\newcommand{\nfg}{n_{\rm fg}}
\newcommand{\nbg}{n_{\rm bg}}
\newcommand{\ncl}{n_{\rm cl}}

\newcommand{\rfg}{r_{\rm fg}}
\newcommand{\rbg}{r_{\rm bg}}
\newcommand{\rcl}{r_{\rm cl}}

\newcommand{\efg}{e^{\rm fg}}
\newcommand{\ebg}{e^{\rm bg}}
\newcommand{\ecl}{e^{\rm cl}}

\newcommand{\hkpc}{h^{-1}\ \kpc}

\shortauthors{Rozo et al.}
\shorttitle{On Stacked Weak Lensing}

\begin{document}
\title{Stacked Weak Lensing Mass Calibration: Estimators, Systematics, and
 Impact on Cosmological Parameter Constraints}
\author{Eduardo Rozo\altaffilmark{1,2}, Hao-Yi Wu\altaffilmark{3}, and 
Fabian Schmidt\altaffilmark{4}}

\altaffiltext{1}{Einstein Fellow, Department of Astronomy \&
 Astrophysics, The University of Chicago, Chicago, IL 60637, USA}
\altaffiltext{2}{Kavli Institute for Cosmological Physics, Chicago, IL
 60637, USA}
\altaffiltext{3}{Kavli Institute for Particle Astrophysics and
 Cosmology, Physics Department, and SLAC National Accelerator
 Laboratory, Stanford University, Stanford, CA 94305, USA}
\altaffiltext{4}{Theoretical Astrophysics, California Institute of
 Technology, M/C 350-17, Pasadena, California 91125, USA}

\begin{abstract}
When extracting the weak lensing shear signal, one may employ either locally normalized or
globally normalized shear estimators.  The former is the standard approach when estimating cluster
masses, while the latter is the more common method among peak finding efforts.  While both
approaches have identical signal-to-noise in the weak lensing limit, it is possible that higher order corrections
or systematics considerations make one estimator preferable over the other.
In this paper, we consider the efficacy of both estimators
within the context of stacked weak lensing mass estimation in the Dark Energy Survey (DES).
We find the two estimators have nearly identical statistical precision, even after including higher order corrections,
but that these corrections must be incorporated into the analysis to avoid observationally relevant biases in the recovered
masses.  We also demonstrate that finite bin-width effects may be significant if not properly accounted 
for, and that the two estimators exhibit different systematics,
particularly with respect to contamination of the source catalog by
foreground galaxies.  
Thus, the two estimators may be employed as a systematics cross-check of each other.
Stacked weak lensing in the DES should allow for the mean
mass of galaxy clusters to be calibrated to $\approx 2\%$ precision (statistical only), which can improve the figure of merit
of the DES cluster abundance experiment 
by a factor of $\sim 3$ relative to the self-calibration expectation.  A companion paper \citep{paperB} investigates how the two
types of estimators considered here impact weak lensing peak finding efforts.
\end{abstract}

\keywords{
cosmology: clusters
}

\section{Introduction}

Upcoming large photometric surveys such as the Dark Energy Survey (DES)\footnote{http://www.darkenergysurvey.org/}, 
Pan-Starrs\footnote{http://pan-starrs.ifa.hawaii.edu/}, 
and the Large Synoptic Survey Telescope (LSST)\footnote{http;//www.lsst.org/} will
find hundreds of thousands of clusters over large fractions of the sky.  These samples hold the potential to be our most powerful tool in understanding 
dark energy \citep{detf06}, and appear  to be a necessary component of any set of observables that wishes to 
distinguish between dark energy and modified gravity approaches for explaining an accelerating universe \citep{shapiroetal10}.  
Indeed, even today cluster abundances provide some of the tightest constraints on the amplitude of the low redshift 
matter power spectrum \citep{mantzetal08,henryetal09,vikhlininetal09,rozoetal10a}, 
and interesting constrains on modifications to gravity \citep{schmidtetal08,rapettietal,lombriseretal}.  
Not surprisingly then, realizing the promise of galaxy clusters as a cosmological probe is of paramount
importance for understanding  the physics driving the current phase of accelerated expansion of the universe.

The most important obstacle that cosmological applications of cluster surveys must overcome is the calibration of mass--observable relations.
That is, the Cold Dark Matter (CDM) paradigm of structure formation allows us to 
predict the abundance of galaxy clusters as a function of mass, whereas empirically we can only recover
the abundance of galaxy clusters as a function of some observable $X$ that correlates with mass.  Consequently, cosmological investigations of
cluster abundances require that we carefully calibrate the probability $P(X|M)$ that a halo of mass $M$ is included in a survey as a cluster with
observable $X$.   This is a problem that is particularly difficult for photometrically selected cluster samples, as there is little theoretical understanding of the
relation between a cluster's galaxy content and its total mass.

Self-calibration is an elegant attempt to overcome this difficulty.  In this approach,
one simply parameterizes $P(X|M)$, and fits for the corresponding parameters relying on clustering information \citep{limahu04,limahu05,hucohn06},
and/or the evolution of abundances with redshift \citep{majumdarmohr04,gladdersetal07}.
One may further enhance such self-calibration techniques by relying on multiple mass tracers \citep{cunhaetal09b}, and these approaches
are expected to be very successful in
improving our understanding of dark energy relative to other dark energy probes \citep{cunhaetal09}.  Nevertheless, it is expected that
careful a priori calibration of the mass--observable relation of galaxy clusters from targeted follow-up observations 
can further enhance the utility of cluster samples over and above
what can be achieved through self-calibration \citep{majumdarmohr03,majumdarmohr04,wuetal10}.

One way of empirically calibrating cluster masses is through cluster weak lensing stacking \citep{sealfonetal06,johnstonetal07b,sheldonetal09,
mandelbaumetal08b,leauthaudetal10,whiteetal10}.
This technique relies on coherently adding the weak lensing signal of galaxy clusters at fixed observable in order
to estimate the mean mass of the stacked galaxy clusters.  Relative to estimating the weak lensing mass of individual clusters,
cluster stacking has the significant advantage of allowing us to detect the lensing signal at significantly lower cluster masses than would
otherwise not be possible.  Moreover, averaging over many halos dramatically reduces the impact of weak lensing projection effects due to non-correlated
structures along the line of sight.  While such measurements do not in any way constrain the {\it scatter} of the observable--mass relation,
the effectiveness of cluster stacking on improving
cosmological constraints from photometrically selected cluster samples is evidenced by the dramatic improvement that this measurement
produced on the cosmological constraints derived from the SDSS maxBCG cluster sample \citep{rozoetal07a,rozoetal10a}.

Here, we forecast the precision with which cluster mass and concentration can be measured in a DES-like photometric survey,
and we discuss a variety of systematics associated with this measurement. 
Specifically, motivated
by the large lensing bias signal expected in the weak lensing power spectrum in the DES \citep{schmidtetal09b}, we revisit the question of
whether lensing bias can have a significant impact on weak lensing mass calibration \citep{mandelbaumetal05b}.   
We also consider finite bin-width correction to density contrast estimates, and finally, we discuss how photometric redshift source selection
can impact the recovered weak lensing signal.

This paper is one of two companion papers;  the second paper considers the problem of identifying weak lensing peaks in large photometric surveys
and the impact of lensing bias on weak lensing peak finding.  An interesting by-product of having performed a simultaneous study of these two distinct
subjects was the realization that in attacking these problems, different approaches for extracting the weak lensing signal are usually taken.   Specifically,
in estimating weak lensing masses one often bins galaxies in annuli, and estimates the mean shear in an annulus by dividing by the number of galaxies
found within the annulus, in other words, adopting a local normalization.  Peak finders, on the other hand, often estimate the shear signal within a
region by simply filtering the shear map which the galaxies sample, without necessarily dividing by the number of galaxies within the filtered region.  
In other words, they employ a global normalization, one that does not depend on the local galaxy density.  
In these papers, we explore how each of these choices of normalization can impact the statistics and systematics of both peak finding and mass estimation.

This paper is organized as follows.  In section 2 we introduce the
weak lensing estimator and calculate the statistical uncertainties of
cluster mass obtained from this estimator.  Various systematic errors
are addressed in section 3.  We study the impact of these mass estimates
on the cosmological parameter constraints in section 4.
Finally, we conclude in section 5.

We adopt a fiducial flat $\Lambda$CDM cosmology with $\Omega_m=0.28$, $\Omega_\Lambda=0.72$, $h=0.7$, $\Delta\zeta=4.54\times 10^{-5}$,
and $n_s=0.96$.  All masses are $M_{200m}$\footnote{i.e. 200 overdensity with respect to the mean matter density.}, 
and distances are {\it physical} distances in either $\kpc$ or $\Mpc$ (as opposed to $\hkpc$
or $\hMpc$).  Finally,  we set the lensing bias parameter $q=1.5$ (see
below for details).


\section{Mass Calibration in Large Optical Surveys Via Cluster Stacking}

\subsection{The Weak Lensing Shear Estimator} 
\label{sec:estimator1}

We consider a shear weak lensing estimator of the form
\be
\DS = \frac{1}{\bar n A} \sum_i \Sc(z_i) e_i 
\ee
Here, the sum is over all galaxies within some annulus of radius $R$,
$e_i$ is the ellipticity of galaxy $i$, $\bar n$ is the mean density
of galaxies, and $A$ is the area of the annulus.  Expert readers will
immediately discern that the estimator $\DS$ is different from the
standard weak lensing shear estimator:
\be
\DS' = \frac{1}{N}\sum_i \Sc(z_i) e_i
\label{eq:std_est}
\ee
where $N$ is the total number of source galaxies within the annulus.

For the next few sections, we will focus primarily on $\DS$ rather
than $\DS'$.
This is primarily for convenience: the fact that the estimator $\DS'$
takes the form $\DS'=x/y$ where $x$ and $y$ are correlated
implies that computing its mean and variance requires some 
additional algebraic gymnastics that we need not worry about when
considering $\DS$.  Consequently, we have opted to illustrate our
discussion with $\DS$ first, and then treat the more complicated
case of $\DS'$.  Section \ref{sec:estimator2} discusses how
the two estimators compare in terms of statistical precision, while
in section \ref{sec:memb_dilution} we demonstrate that the two estimators
have significantly different systematics with respect to foreground
contamination of the source galaxy population.  

We wish to estimate the mean and variance of $\DS$.  In the interest of simplicity, we assume all sources are at the same
redshift $z$ and set $\Sigma_c(z_i)=\bar \Sigma_c$.
It is also useful to rewrite our estimator as follows:
first, we pixelize the sky behind the lens in pixels of area $\Delta\Omega$, 
and define $N_i$ as the number of galaxies in pixel $i$.  One might naively expect
\be
N_i=\bar n \Delta\Omega (1+\delta_i),
\ee
where the $\delta_i$ is the overdensity of matter in pixel $i$.
In practice, however, lensing modifies the observed source galaxy density such that \citep{schmidtetal09}
\be
N_i = \bar n \Delta\Omega \mu_i^{q/2}(1+\delta_i)
\ee
where $\mu_i$ is the magnification evaluated at pixel $i$, and $q$ is a number that characterizes how the source density changes
due to gravitational lensing.  For a DES-like survey, we expect $q\approx 1-2$ \citep{schmidtetal09}.
Defining the filter function $W_i$ such
that $W_i=1$ only when a pixel is within the annulus of interest, and assuming the pixels are small enough that $N_i$ is always either zero or one,
we rewrite $\DS$ as
\be
\DS = \frac{\bar \Sc}{A}\sum_i \Delta \Omega\  \mu_i^{q/2}(1+\delta_i) W_i e_i 
\ee
where the sum is now over all pixels.  This is the equation we will use to derive the expectation value
and variance of $\DS$.


\subsection{Mean and Variance of $\DS$}

The expectation value of $\DS$ follows from direct substitution of the expectation values of $e_i$ and $\delta_i$.  Specifically, 
\bea
\avg{e_i} & = & g_i \label{eq:mean_ellip}\\
\avg{\delta_i} & = & 0
\eea
where $g$ is the reduced shear, $g=\gamma/(1-\kappa)$.  We assume now that we work in the thin annulus approximation
so that the variation of $\mu$ and $g$ within an
annulus is negligible (though see also section \ref{sec:finite_width}), and find 
\be
\avg{\DS} = \bar \Sc \mu^{q/2} g \approx \Sc \gamma 
\ee
where all lensing quantities are to be evaluated at the annulus radius
of interest.  The approximate equality is obtained by expanding
to leading order in the lensing quantities.

To compute the variance, we assume source galaxies are unclustered, so that $N_i$ is Poisson.  We have then
\be
\Cov(N_i,N_j) = \delta_{ij}\Var(N_i) = \delta_{ij} \avg{N_i} = \delta_{ij} \mu^{q/2}\bar n \Delta\Omega,
\ee  
and therefore
\be
\avg{\delta_i\delta_j} = \frac{\Cov(N_i,N_j)}{\avg{N_i}\avg{N_j}} = \delta_{ij}\frac{1}{\mu^{q/2} \bar n \Delta\Omega}.
\label{eq:poisson}
\ee
In appendix \ref{app:src_clustering}, we consider the additional contribution to the noise due to clustering of the source population
and we demonstrate that it is sub-dominant at high masses, and negligible at low masses, justifying the assumption above.
We further assume the variance in the ellipticity of galaxies is dominated by shape noise,
\be
\avg{e_ie_j} = g_i g_j + \delta_{ij}\frac{\sigma_e^2}{2},
\label{eq:var_ellip}
\ee
the factor of two coming from the fact that there are two independent ellipticity components and we use only one of them.
For quantitative purposes, we adopt $\sigma_e=0.3$ as our fiducial value for the amplitude of the shape noise term.
Direct substitution results in 
\be
\Var(\DS) =  \bar \Sc^2 \frac{\mu^{q/2}}{\bar n A} \left( g^2 + \frac{1}{2}\sigma_e^2 \right) \approx \bar \Sc^2 \frac{\sigma_e^2}{2\bar n A},
\label{eq:dserr}
\ee
where again the approximate equality gives the leading order term in a series expansion of the lensing quantities.

The signal-to-noise of our estimator $\DS$ is 
\be
(S/N) = \mu^{q/4}\frac{(\bar n A)^{1/2} g}{(g^2+\sigma_e^2/2)^{1/2}} \approx \frac{(\bar n A)^{1/2} \gamma}{(\sigma_e^2/2)^{1/2}}.
\label{eq:sn_ds}
\ee
In a DES-like survey, we expect $q\approx 1-2$, in which case the lensing bias correction $\mu^{q/4}$ results in a modest increase of the signal-to-noise
of our measurement.


\subsection{Survey Assumptions, Fiducial Model, and the Fisher Matrix}
\label{sec:Fisher}

We wish to estimate the precision with which stacked shear weak lensing experiments can constrain halo mass and concentration
in a DES-like survey.  We adopt a survey area $\Omega=5,000\ \deg^2$, 
a fiducial galaxy density $\bar n= 10\ \mbox{galaxies}/\mbox{arcmin}^2$,
and a source redshift distribution
\be
f(z) \propto z^m \exp\left[ -(z/z_*)^\beta \right]
\label{eq:fz}
\ee
with $z_*= 0.5$, $m=2$, and $\beta=1.4$, as appropriate for a DES-like
survey.     
The fraction
of galaxies above a given redshift $z$ is given by
\be
F_{bg}(z) = \frac{\int_z^\infty dz'\ f(z')}{\int_0^\infty dz'\ f(z')}.
\ee
Given a lens redshift $z_L$, the (approximate) effective source density is simply $\bar n F_{bg}(z_L)$.
For a characteristic lens redshift $z_L\approx 0.5$, the corresponding source density
is $\approx 8\ \mbox{galaxies}/\mbox{deg}^2$.

We further assume clusters are binned in narrow redshift slices $z=\bar z \pm 0.05$,
and are logarithmically binned in mass in bins of width $\pm \Delta \log_{10} M=0.1$, corresponding to
5 bins per decade in mass.
The effective source density for a given mass and redshift bin is therefore
\be
\bar n_{\rm eff} = 2 \bar n F_{bg} (z) \frac{dn}{d\ln M} \Delta \ln M \Delta V
\label{eq:neff}
\ee
where the factor of 2 arises from the total width of the mass bin $2\Delta\ln M$, $\Delta V$ is the survey volume enclosed by
the redshift slice $z\pm \Delta z$, 
\be
\Delta V = (1+z)^2 D_A^2(z) \Omega c \Hinv(z) 2\Delta z,
\ee 
$D_A$ being the angular diameter distance, 
and $F_{bg}(z)$ is the fraction of source galaxies with redshift higher than $z$.
When estimating all lensing properties, we will further assume that all source galaxies behind the clusters reside at a single
source redshift $z_s$ equal to the mean source redshift of the sources behind the cluster.
The mass function $dn/d\ln M$ is computed using the \citet{tinkeretal08} mass function in our fiducial cosmology.
As mentioned in the introduction, for the purposes of computing the impact of lensing bias we always assume
$q=1.5$ \citep[][]{schmidtetal09}.

Note that we have not taken into account the effects of finite mass bins
here:  since the scale radius of halos evolves with mass, the stacked
shear profile within $[\ln M-\Delta\ln M;\ln M+\Delta\ln M]$ is not equal to 
the profile of an NFW halo with mass $M$.  While this needs to be taken into account
when fitting actual data, it is not of direct relevance to our Fisher forecast,
so we neglect this effect here.  Furthermore, it should be noted that, 
observationally, cluster stacks are made by binning in an observable $X$, 
whereas the problem  we have laid out here assumes that the clusters are 
binned in mass.  
While this obviously affects the distribution of cluster masses within a stack, we do not expect
the precision of the corresponding measurement to be particularly sensitive to said distribution.
Indeed, roughly speaking the total uncertainty in the mass per cluster is just the sum in quadrature
of the measurement error with the intrinsic scatter.  For a typical cluster, the latter is smaller than the former,
especially at low masses where most of our cluster sample resides, and therefore our approximation is valid.
Given our expectations and the fact that 
our Fisher Matrix forecast ought to be interpreted as a rough estimate of the precision of these type
of analysis, we have ignored this (mass proxy-dependent) complication.  

Assuming sources are uncorrelated and that the redshift slices are narrow enough that halos are non-overlapping, 
the estimators $\DS$ for different radial bins are uncorrelated.  In this limit,
the Fisher matrix for a weak lensing shear experiment simplifies 
to\footnote{In the interest of simplicity, in equation \ref{eq:fisher} we have neglected the modicum amount of 
information in the small dependence
of $\Var(\DS)$ on the model parameters $\bm{p}$.  We do not expect our results to be sensitive to this detail.  Moreover,
in practice we expect the covariance matrix for the measurements will be estimated using a jackknife method, which will
erase any information in the variation of the covariance matrix with $\bm{p}$. }
\be
F_{ab} =  \sum_\alpha \frac{1}{\Var(\Delta\Sigma_\alpha)} \frac{\partial \avg{\DS_\alpha}}{\partial p_a} \frac{\partial \avg{\DS_\alpha}}{\partial p_b}
\label{eq:fisher}
\ee
where the sum is over all radial bins $\alpha$ and  $\bm{p}$ is the vector of parameters of interest.   Throughout, we assume logarithmic
radial binning with bins of width $\pm \Delta \log_{10} R = 0.02$.  Moreover, we will only add bins over
the radial range $0.1\ \Mpc$ to $2\ \Mpc$.  We will further impose the condition that the cluster magnification must be smaller than five (i.e. $|\mu|\leq 5$)
for a radial bin to be included in our computation.
This ensures that for those systems where the Einstein radius extends past the $0.1\ \Mpc$ minimum radius, we do not include information from sources that are strongly lensed.  Our results are robust to making our bins narrower, but do depend in detail on the radial range assumed.
For a discussion, see section \ref{sec:sensitivity}.   

We model the mass distribution of cluster stacks as a Navarro, Frenk,
and White \citep[NFW,][]{nfw96} profile.  For simplicity, we have held
the concentration parameter of our clusters fixed to $c=5$.  Changes
in concentration within a stack do not qualitatively change any of our 
results, and
have only a modest impact on the quantitative results.  All of our
conclusions are robust to the choice of concentration parameter.  We
compute the convergence and shear of NFW profiles using the formulae
in \citet{wrightbrainerd00} \citep[see also][and appendix of \cite{paperB}]{bartelmann96}.

\subsection{Results}
\label{sec:staterr}


\begin{figure}
\epsscale{1.2}
\plotone{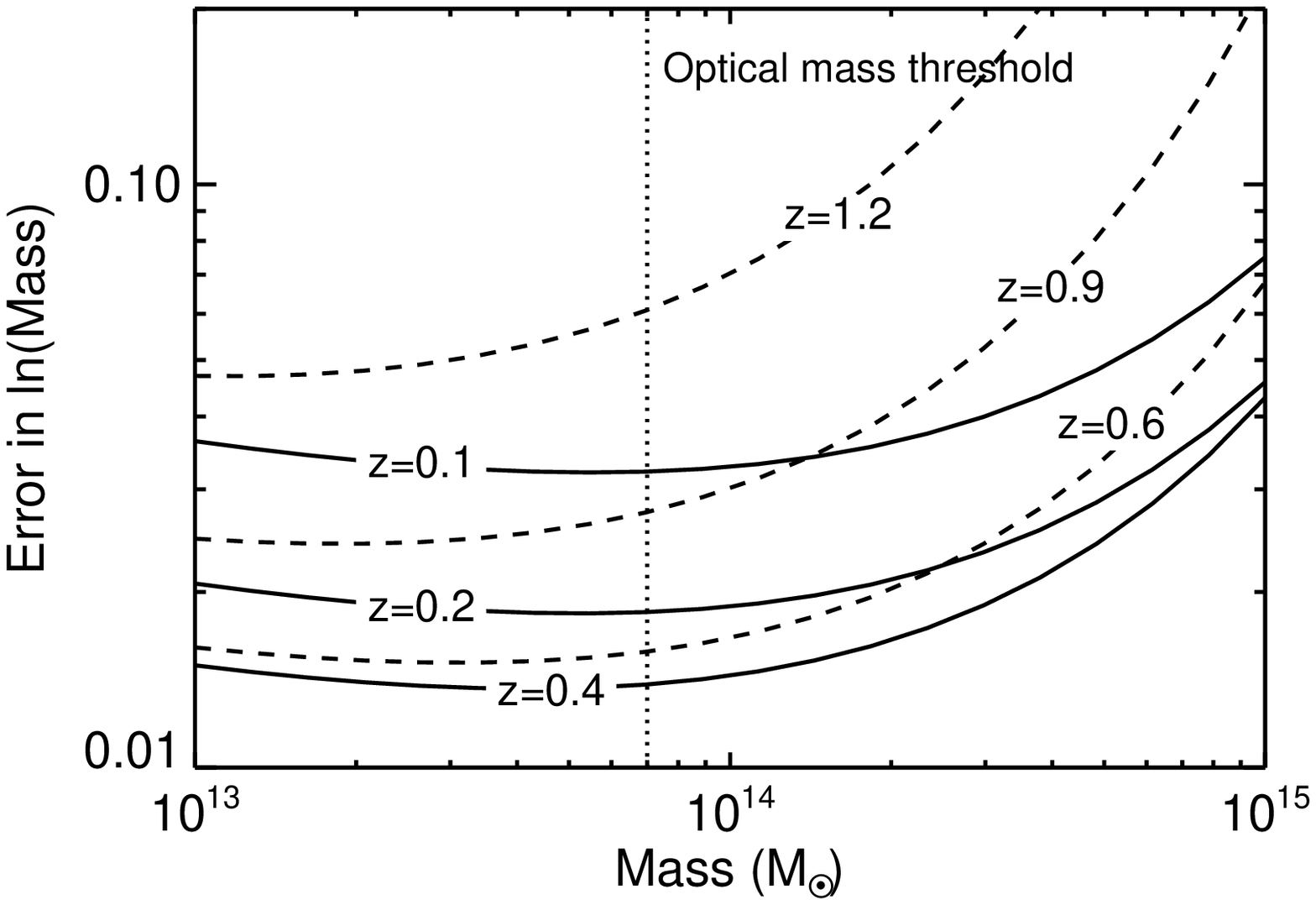}
\plotone{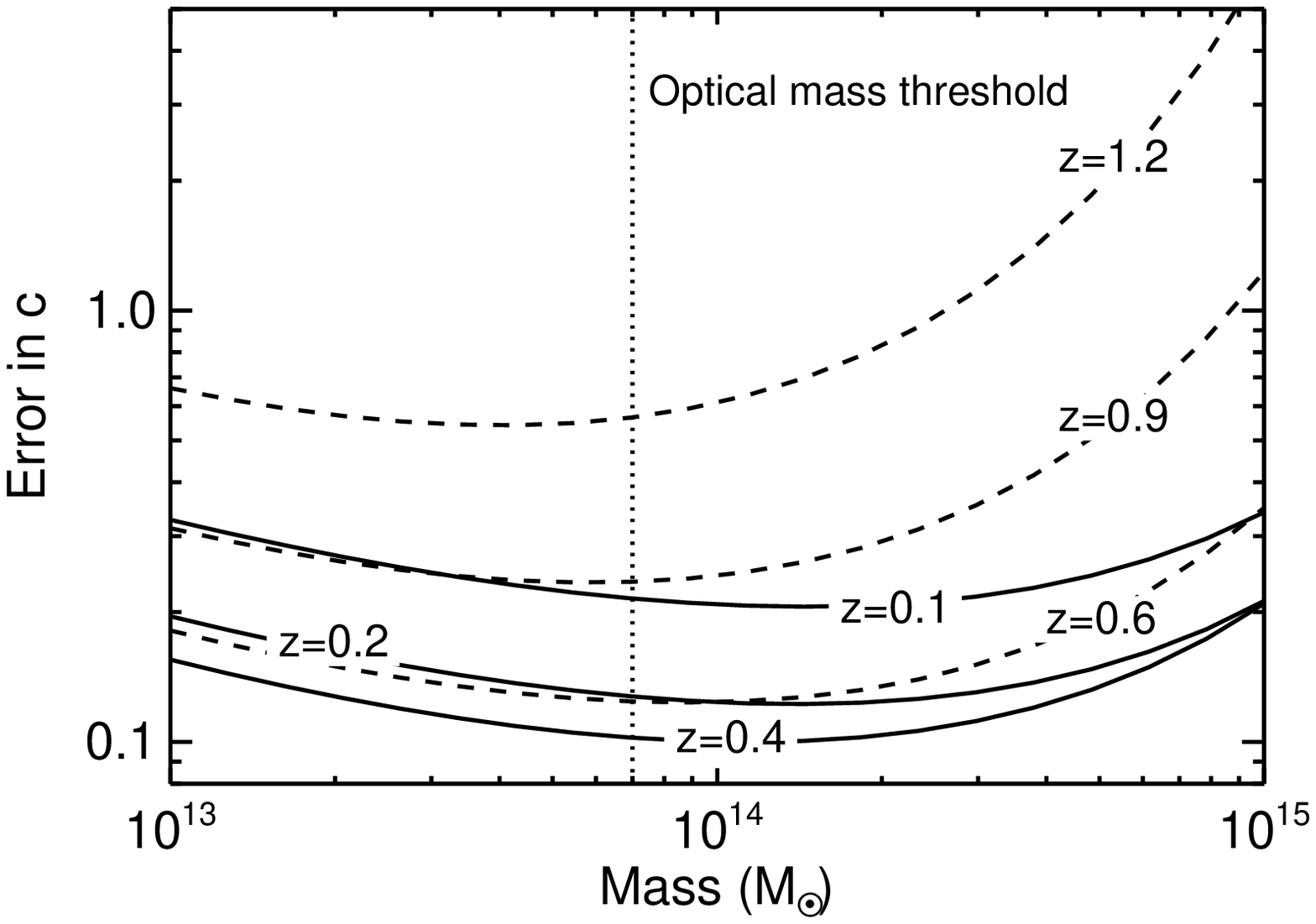}
\caption{The predicted $1\sigma$ {\it statistical} error in the log-mass (top) and the concentration (bottom)
of galaxy clusters in a DES-like survey as measured through cluster stacking.  The solid lines illustrate the trends with redshift for $z\leq 0.4$, while
the dashed lines illustrate the trends with redshift for $z\ge 0.4$.  From top to bottom (in either plot) the lens redshifts of the solid curves
are $z_{lens}=$0.1, 0.2, 0.4, whereas the redshift of the dashed curves is, from bottom to top, $z_{lens}=$0.6, 0.9, and 1.2.
The typical precision with which the mean mass of galaxy clusters may be estimated at moderate redshifts is roughly $2\%$.
}
\label{fig:errors}
\end{figure} 


Figure \ref{fig:errors} shows the predicted statistical uncertainty in the weak lensing mass and concentration of our cluster stacks
as a function of the mean cluster mass of the stack.  Different lines correspond to different lens redshift.  From top to bottom, the solid lines assume
lens redshifts $z_{lens}=0.1$, $0.2$, and $0.4$.  From bottom to top, the dashed lines assume $z_{lens}=0.6$, $0.9$, and $1.2$.
The vertical dotted line is a rough estimate of the expected mass threshold for optical selection.
Qualitatively, the precision of weak lensing measurements increases from $z=0$ to $z\approx 0.2$, reflecting the
increasing number of lenses due to increased survey volume, as well as the improvement on the lensing efficiency of
the lenses.
Between $z\approx 0.2$ and $z\approx 0.6$, the precision of the weak lensing measurements is roughly redshift independent: even though survey volume
continues to increase, it is now offset by a diminishing source density.  For redshifts $z\gtrsim 0.6$, both the lensing efficiency and the effective source density
decrease quickly with increasing lens redshift, so weak lensing masses begin to worsen.
We also note that at all redshifts there is a mass scale at which the errors blow up, reflecting the exponential drop off in the halo mass function.
The typical precision with which the mean mass of galaxy clusters may be estimated at moderate redshifts is roughly $2\%$.


\subsection{On the Choice of Estimator}
\label{sec:estimator2}

In section \ref{sec:estimator1} we defined the estimator $\DS$
differently from the more commonly used weak lensing shear
estimator:
\be
\DS' = \frac{1}{N}\sum_i \Sc(z_i) e_i \nonumber
\ee
Pixelizing the sky, the estimator $\DS'$ can be written as
\be
\DS' = \bar \Sc \frac{\sum_i \Delta\Omega \mu_i^{q/2}(1+\delta_i)W_i e_i}{\sum_i \Delta\Omega \mu_i^{q/2}(1+\delta_i)W_i}
\label{eq:DS1}
\ee
where the sum is now over all pixels.  

A difficulty with $\DS'$ is now immediately apparent.  We see that $\DS'$ takes the form $\DS'=x/y$, and
in general one has $\avg{\DS'} = \avg{x/y} \neq \avg{x}/\avg{y}$.  One way to address this problem is to write
$y=\avg{y}+\delta y$, and expand the denominator
in a power series using the binomial expansion.  This results in a power series expansion of all quantities in terms 
of $1/(\bar n A)$ (see Appendix \ref{app:dsprime}).  
Doing so up to second order in $\delta$ to compute the mean of $\DS'$ we find
\be
\avg{\DS'} = \bar \Sc \frac{\avg{\mu^{q/2}g}}{\avg{\mu^{q/2}}}.
\ee
The quantities in the angular brackets in the right hand side of the equation are to be averaged over the radial bin of interest.  
Adopting the thin annulus approximation, 
we can pull out the $\mu_i^{q/2}$ term out of the sums in equation \ref{eq:DS1} to obtain
\be
\DS' = \bar \Sc \frac{\sum_i \Delta\Omega (1+\delta_i) e_i}{\sum_i \Delta\Omega (1+\delta_i)}.
\ee
In this limit, the magnification term drops entirely out of the equations, and one can easily solve for the mean and variance of $\DS'$
using the power series approach advocated earlier.  We find (see Appendix \ref{app:dsprime})
\bea
\avg{\DS'} & = & \bar \Sc g  \label{eq:avgdsprime} \\
\Var(\DS') & = & \bar \Sc^2 \frac{1}{\mu^{q/2} \bar n A} \frac{\sigma_e^2}{2} \label{eq:vardsprime}. 
\eea
Note the magnification still appears in the variance since the effective source density is $\mu^{q/2}\bar n$.

Using the above equations,
we repeat the Fisher matrix experiment that we carried out for $\DS$.
We find that these two estimators recover the mean cluster mass of a stack
with nearly identical precision, with 
$|\sigma_M-\sigma_{M'}| \lesssim 0.2$\% where $\sigma_M$ and $\sigma_{M'}$ are the 
forecasted uncertainties in $\avg{\ln M}$ for a stack using the estimators $\DS$ and $\DS'$ respectively.

Before we end, we would like to put a word of warning concerning the validity of our results for the estimator $\DS'$.
Our derivation in this section deals with the covariance between the numerator and denominator of $\DS'$ through
a power series expansion, where the expansion parameter is $1/\bar n A$, the expected number of galaxies
within the annulus of interest.  Consequently, for our analysis to be applicable one must choose 
radial bins that are large enough for $\bar n A \gg 1$.  On the other hand, we are also using the thin annulus 
approximation, so $A$ cannot be arbitrarily large.  While we expect to be able to simultaneously satisfy both of these constraints in stacked
weak lensing analysis, this is generally not possible for individual clusters.  For instance, in our fiducial model,
a relatively broad radial bin $R\in[0.2\ \Mpc,0.4\ \Mpc]$ would only contain $\approx 25$ galaxies, so the
higher-order terms in the $1/\bar n A$ power series expansion can in principle introduce $\approx 5\%$ level corrections
to the signal in the inner most radial bins.  
Here, we do not address this additional difficulty since our primary interest is stacked weak lensing.
The finite width bin corrections that we treat below concern corrections to the leading order term which occur even
when $\bar n A \gg 1$.


\section{Systematics}

\subsection{Lensing Bias Corrections}
\label{sec:important?}

By lensing bias corrections we refer to the terms that scale as $\mu^{q/N}$ for some value of $N$ in both the mean
and variance of our shear estimators.  As discussed in section \ref{sec:estimator2}, in the thin annulus approximation 
the mean of $\DS'$ is independent of these lensing bias terms, so we expect lensing bias corrections to be 
small.\footnote{In principle, the fact that lensing bias does impact the covariance matrix of $\DS'$ could lead
to biasing in halo mass and concentration.  In practice, however, the error bars employed when relying on $\DS'$
are based on the Poisson variance of the {\it observed} number density, which
includes the appropriate lensing bias.  Thus, one does not expect $\DS'$ to be affected by lensing bias.}
The same is not true
of $\DS$, and therefore ignoring these corrections can potentially lead to systematic biases.  That said,
it is worth emphasizing that incorporating these corrections into the analysis is not difficult, so having to include
such corrections is not any sort of limiting systematic.  Our goal here is simply to determine whether 
doing so is a necessary step in the analysis of future data sets.


\begin{figure}[t]
\epsscale{1.2}
\plotone{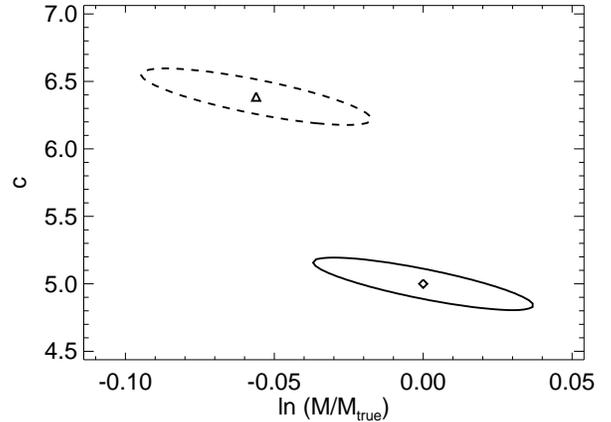}
\caption{$68\%$ likelihood contours obtained when measuring the mass and concentration of a halo stack of mass $M=5\times 10^{14}\ \msun$
at redshift $z=0.3$ using a DES-like survey.  The solid contour includes the corrections for lensing bias, while the dashed line ignores these corrections.
The best fit value for the former case coincides with the input parameters, marked by a diamond, while ignoring these corrections leads to biased
expectation values, shown here with a triangle.  This behavior is generic (see Figure \ref{fig:bias} for details).
}
\label{fig:ellipse}
\end{figure} 


To address this question, we use the Fisher matrix formalism set out in the appendix of \citet{wuetal08} \citep[see also][]{hutererlinder07}
to estimate the mass
and concentration parameters that would be recovered from the data if lensing bias corrections are ignored,
and then compare those results to those derived when properly including the lensing bias terms in the analysis.
The basic idea is illustrated in Figure \ref{fig:ellipse}.  The solid and
dashed ellipses represent the $68\%$ confidence contours when including (solid) or ignoring (dashed) lensing bias corrections
for a halo stack at $z=0.3$ and $M=5\times 10^{14}\ \msun$.
The true halo mass and concentration is marked by the diamond, while the triangle marks the best fit halo mass and concentration
obtained when ignoring lensing bias corrections.
We can
see that the best fit values when ignoring lensing bias effects  are well beyond the $68\%$ error ellipse of the experiment,
demonstrating that lensing bias is a significant correction when using the $\DS$ estimator.

\begin{figure}[t]
\epsscale{1.2}
\plotone{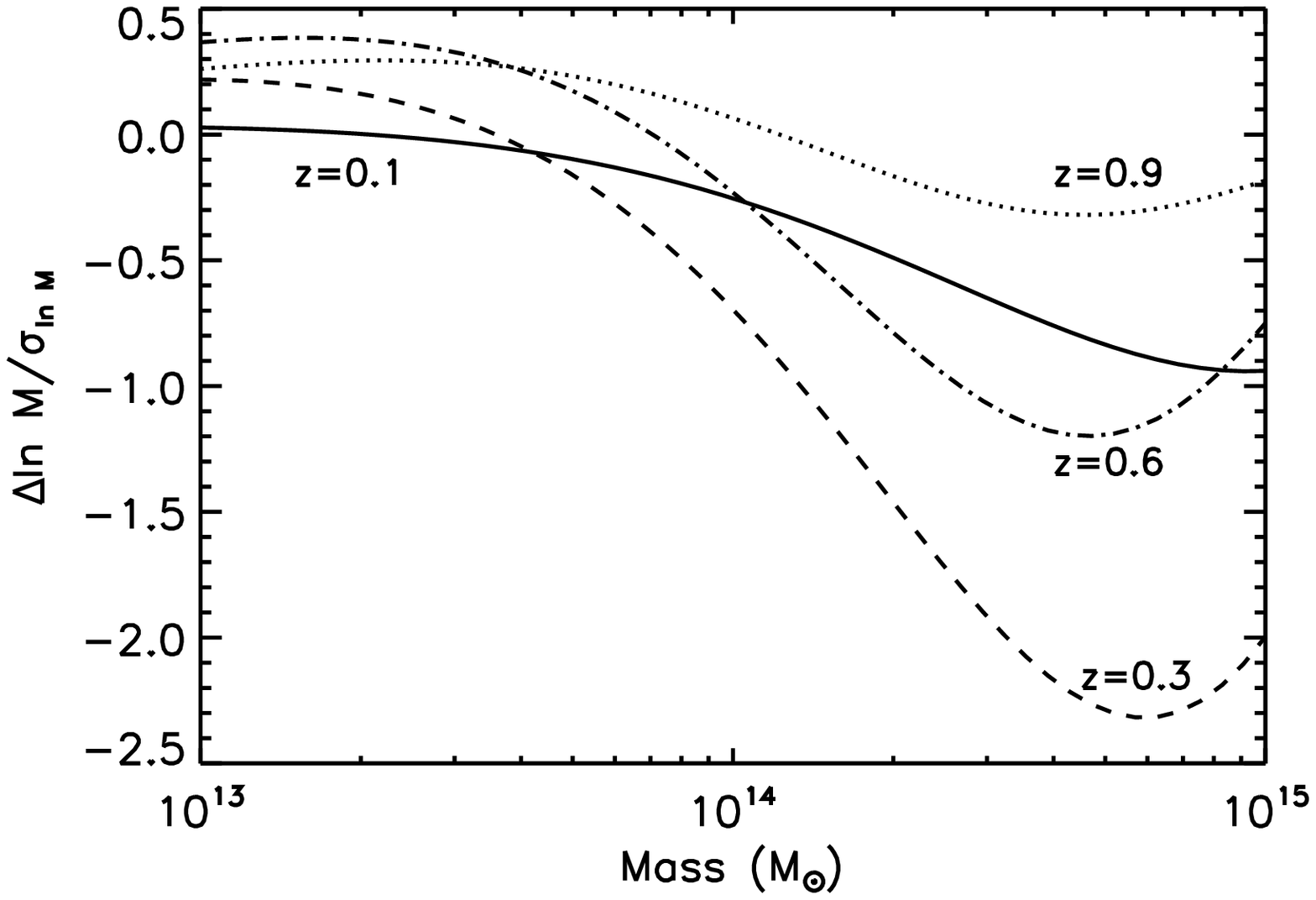}
\plotone{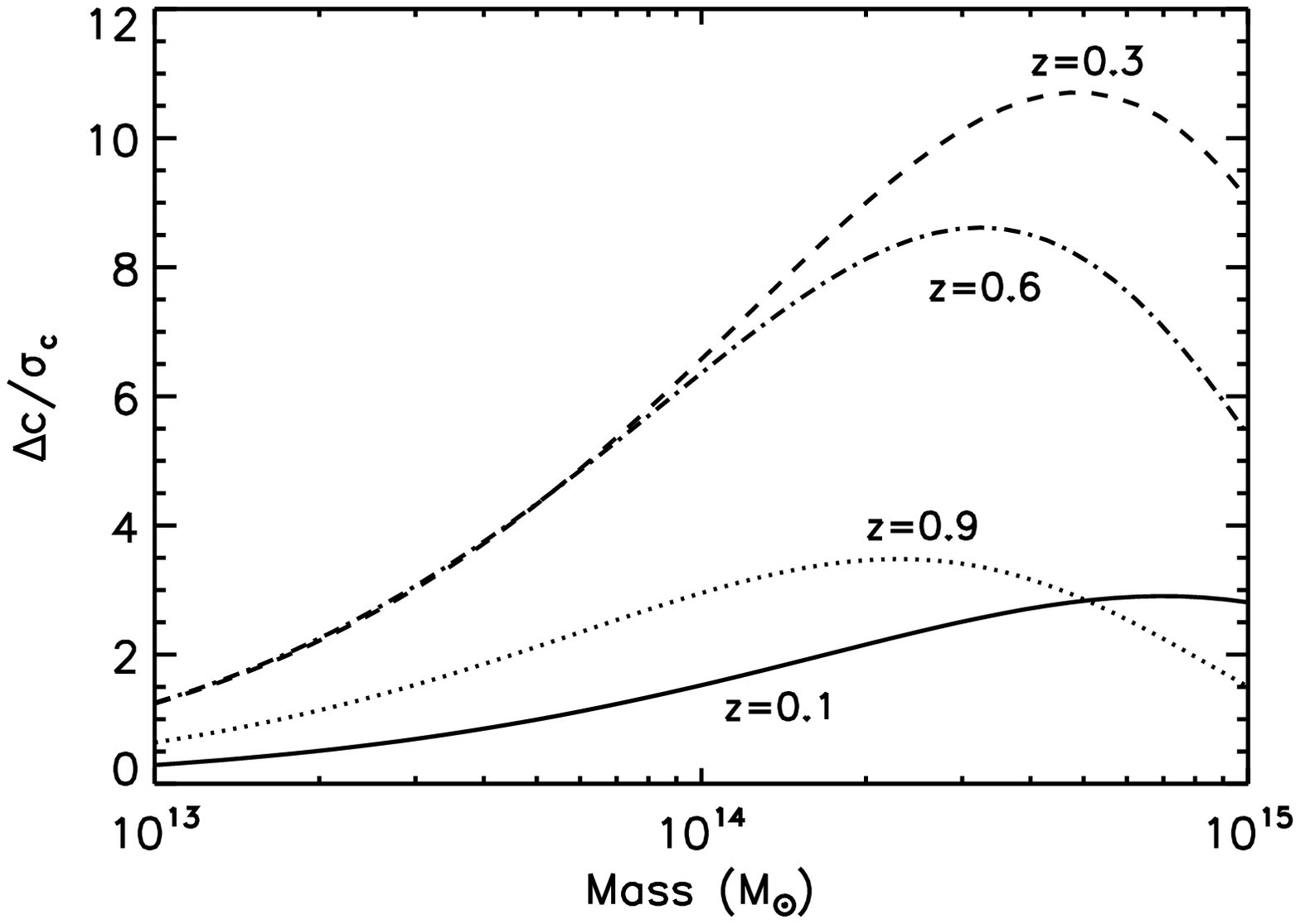}
\caption{Ratio of the systematic bias in the log-mass $\Delta\ln M$ or concentration $\Delta c$ incurred when ignoring lensing bias and reduced shear errors, 
relative to the statistical uncertainty $\sigma_{\ln M}$ or $\sigma_c$ in these quantities (see Figure \ref{fig:errors}).  
}
\label{fig:bias}
\end{figure}

The extent to which lensing bias is important relative to statistical uncertainties depends on both halo mass and redshift.
In particular, lensing bias is more important at high masses, and at redshifts $z\approx 0.2-0.5$, for which lensing efficiency
is high and statistical errors are small due to high source densities.   This is illustrated in Figure \ref{fig:bias}, where we show
the ratio between the systematic bias in mass and concentration incurred by ignoring lensing bias effects, relative to the corresponding
statistical uncertainty.
We see that our projected bias in the halo mass can be twice as large as the corresponding statistical uncertainty,
while the corresponding errors in concentration can be even larger than 10 times the statistical error.  Thus,
inclusion of lensing bias and reduced shear corrections are 
necessary for weak lensing mass calibrations in a DES-like survey.  


One aspect of our results that is very surprising is that lensing bias can lead to systematic {\it underestimates} of the mass, despite the fact that lensing
bias leads to an apparent increase of the lensing signal.  The reason we find that masses are underestimated when ignoring lensing bias is due
to the anti-correlation between halo mass and concentration.  Since lensing bias dramatically boosts the recovered concentration, the recovered mass
must necessarily go down to compensate.  This statement, however, is very much dependent on the range of radii considered in the calculation.
In particular, as one moves out the maximum radius employed in the analysis the anti-correlation between mass and concentration decreases, and lensing
bias begins to produce positively biased halo masses (see also section \ref{sec:sensitivity}).

Finally, our results may appear to contradict those in \cite{paperB}, who find lensing bias results in an overestimate 
of weak lensing masses despite the fact that they estimate masses using filters of comparable radius to those employed here.
The reason for this apparent discrepancy is that in this analysis we have fit for both halo mass and concentration, whereas in the
\cite{paperB} analysis, the concentration parameter is held fixed at its fiducial value.  In that case, the apparent increase in the matter density 
in the inner regions of a cluster due to lensing bias necessarily goes into halo mass rather than concentration, leading to a positive mass bias.
In practice, because cluster stacks have abundant signal, one fits for both
mass and concentration, so the observationally relevant case for cluster stacks is that considered here.  For individual weak lensing
mass estimates in the low signal-to-noise regime, the results of \cite{paperB} are the relevant ones.


\subsection{The Relative Information Content of Different Scales}
\label{sec:sensitivity}

In light of the results from the previous section
it is worth investigating the contribution to the Fisher matrix from each individual radial bin as a function of radius.
The information contributed by radial bin $R_\alpha$ is
\be
F_{ab}(R_\alpha) =  \frac{1}{\Var(\Delta\Sigma_\alpha)} \frac{\partial \avg{\DS_\alpha}}{\partial p_a} \frac{\partial \avg{\DS_\alpha}}{\partial p_b}.
\ee
Of course, this quantity explicitly depends on the width of the radial bins employed.  We therefore define the {\it relative sensitivity} $s_{ab}$ via
\be
s_{ab}(R_\alpha) = \frac{ F_{ab}(R_\alpha) }{\max\{ F_{ab}(R_\alpha) \} } 
\ee
assuming logarithmic radial bins.  Note that, by construction, $s_{ab}$ is very nearly 
independent of the assumed width of the radial bins and the assumed source density.
Of particular interest to us are the mass sensitivity $s_{mm}$ and concentration sensitivity $s_{cc}$, which characterize the relative mass and concentration 
information content of various scales.


\begin{figure}[t]
\epsscale{1.2}
\plotone{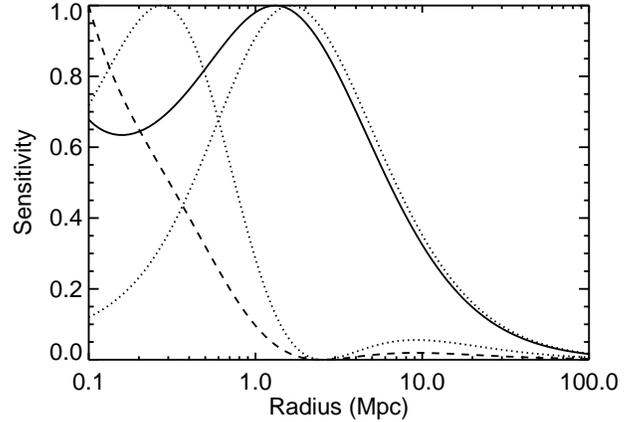}
\caption{Mass (solid) and concentration (dashed) sensitivity as a function of radius for a $10^{15}\ \msun$ halo at redshift $z=0.6$.  
The sensitivity functions characterize the relative information content of various scales.  Dotted curves are the two sensitivity
functions we obtain when neglecting lensing bias corrections.  Not surprisingly, the scales that dominate the concentration and
mass information content are a few hundred kpc and a few Mpc respectively.  In both cases, 
lensing bias shifts the sensitivity functions towards smaller scales, as we would expect.
}
\label{fig:sensitivity}
\end{figure} 


Figure \ref{fig:sensitivity} illustrates the mass (solid) and concentration (dashed) sensitivity functions for a $10^{15}\ \msun$ halo at redshift $z=0.6$.  
Not surprisingly, the mass information content is dominated by scales of order a few Mpc, while the concentration information content is dominated
by scales of order a few hundred kpc.  The large overlap of the two sensitivity functions results in the anti-correlation between halo mass and concentration
noted earlier.  The dotted lines in the figure correspond to the sensitivity functions in the absence of lensing bias effects (i.e. setting $q=0$).  We see
that lensing bias shifts the sensitivity functions to smaller scales, effectively up-weighting them, as we would expect.
The corresponding sensitivity functions for the estimator $\DS'$ are essentially identical to those of $\DS$ in the absence
of lensing bias corrections.

In light of Figure \ref{fig:sensitivity} it is easy to understand why our results were sensitive to the adopted radial range.  It is evident from the figure that
both the minimum and maximum radius will impact our results.  The minimum radius chosen will directly affect the precision of the concentration
measurement, which in turn propagates to the cluster mass via the mass--concentration correlation.  Likewise, it is easy to see why our results are
sensitive to the maximum radius employed: scales as large as tens of Mpc still contain non-negligible amounts of information pertaining to
the halo mass.  Of course, in practice such scales show deviations from a simple projected NFW profile (\cite{Hoekstra02}), both because of possible deviations 
from NFW in the 1-halo term,
and because of the appearance of a 2-halo term.   We defer an investigation of this problem to future work.
Here, we simply note that we expect that there is significant additional information in the weak lensing signal up to scales as large as several tens of Mpc.


\subsection{The Thin Annulus Approximation}
\label{sec:finite_width}

When performing our forecasts for both $\DS$ and $\DS'$ we adopted the thin annulus approximation, in which
one ignores variations of the convergence, shear, and magnification fields within an annulus.
When using finite radial bin widths, however, there is a slight ambiguity on what radius should one employ
when performing the thin annulus approximation.   Two obvious choices are the central radius $R_c$
where the radial bins are defined by $\ln R_c \pm \Delta \ln R$, and the mean radius of the annulus
\be
\avg{R} = \frac{2}{3}\frac{R_{max}^3-R_{min}^3}{R_{max}^2-R_{min}^2}.
\ee
The question then becomes, how thin must an annulus be for the thin annulus approximation to hold
for either of these two choices of radii?

To address this question, we have again relied on a Fisher matrix
approach to estimate the systematic error in the log-mass of a galaxy cluster when the mass is estimated using the thin-annulus approximation,
as a function of the width of the annulus used to estimate $\DS$ or $\DS'$.    For specificity, we will focus here on $\DS'$, but very similar
conclusions hold for $\DS$.  We emphasize that, just as with lensing bias corrections, one can in principle just include these finite bin width
effects in the fits.  The analysis described here simply addresses whether it is necessary to do so.


\begin{figure}[t]
\epsscale{1.2}
\plotone{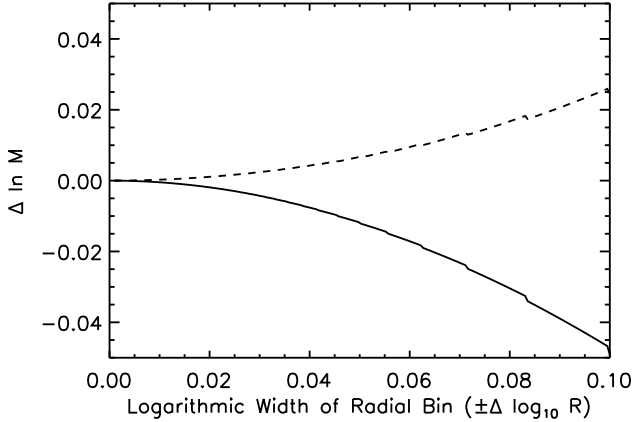}
\caption{Bias in the recovered weak lensing masses using the thin-annulus approximation for predicting the expectation value of $\DS'$
as a function of the width of the annulus used to estimate the shear profile.  We assume $\DS'$ is estimated in annuli of width $\Delta$
defined via $[\ln R_c-\Delta,\ln R_c+\Delta]$.  The solid line assumes the thin annulus approximation
at $R=R_c$, while the dashed line assumes the thin annulus approximation at the mean radius of the annulus $\avg{R}$.
This bias is nearly independent of halo mass, concentration,
redshift, and source density.  
}
\label{fig:ds_prime_mass_bias}
\end{figure} 


Our results are shown in Figure \ref{fig:ds_prime_mass_bias}.
The largest bin width we consider is roughly five bins per decade in radius, corresponding to $\pm \Delta \log_{10} R = 0.1$. For such a bin width,
there is a negative mass bias of $\approx 5\%$ assuming the thin annulus approximation at $R=R_c$ (solid line), 
and a $\approx 2\%$ positive bias assuming
$R=\avg{R}$ (dashed line).  This bias is nearly independent of halo mass, concentration, redshift, or source galaxy density.  Thus, it is important
to account for finite bin width effects when estimating cluster masses, particularly for the purposes of calibrating cluster scaling 
relations.\footnote{The small ``steps'' that appear in Figure \ref{fig:ds_prime_mass_bias}
are real.  To estimate cluster masses, we use the scales $0.1\ \Mpc < R < 2\ \Mpc$.  This means that when we vary the bin width,
there will be discrete jumps in the number of bins, which leads to the ``stepping'' observed in the plot.}

The origin of the the above bias is simple to understand: consider first the case $R=R_c$.  It is evident that in this case our annulus
contains more area beyond $R=R_c$ than below $R=R_c$.  Consequently, we expect a negative bias, since the outer region of the annulus
is being up-weighted relative to the inner region.  Setting $R=\avg{R}$ evidently overcompensates for this effect.  While one can in principle imagine
finding the specific value of $R$ for which there is no bias, this value will depend in detail on cosmology, source galaxy density (due to magnification
bias corrections), etc., so it is simplest to just take into account the finite bin-width correction exactly.
Alternatively, when performing stacked weak lensing measurements one may adopt very narrow radial bins, in which case these corrections
become negligible.


\subsection{Foreground and Cluster Member Dilution}
\label{sec:memb_dilution}

One important systematic when estimating the mass of galaxy clusters through weak lensing is the effect known as member dilution (\cite{Bernardeau98,MedezinskiEtal07}).  That is, if cluster members or other
foreground galaxies are included in the source population, they will reduce the observed shear and cause us to underestimate the cluster mass.  Let us treat this problem
explicitly: the total (angular) galaxy density around a cluster at redshift $z$ can be written as a sum
\be
n = \nbg + \nfg + \ncl 
\ee
where $\nbg$ is the background galaxy density field, $\nfg$ is the foreground galaxy density, and $\ncl$ is the density of cluster galaxies.  When estimating
weak lensing masses, one adopts photometric redshift cuts to preferentially select background galaxies.  Let $\rbg$, $\rfg$, and $\rcl$ be the probabilities
of passing the photometric redshift cuts for a galaxy in the background, 
foreground, and at the lens redshift, respectively.  A perfect photometric 
redshift selection would result in $\rfg=0$, $\rcl=0$, and $\rbg=1$, so the 
value of these parameters characterizes the actual redshift distribution
of the source sample in the presence of imperfect photometric redshift selection.  
We wish to consider the impact of uncertainties in these parameters on $\DS$ and $\DS'$.  For simplicity, we will work in the thin annulus limit 
throughout.  

We begin by considering $\DS$, which we now write as
\bea
\Delta\Sigma & = & \frac{\bar \Sc}{\bar n A F} \sum_i \Delta\Omega 
\left[ \rbg \nbg \ebg_i + \rfg \nfg \efg_i  \right] \nn \\
	& & + \frac{\bar \Sc}{\bar n A F} \sum_i \Delta \Omega \ncl \ecl_i 
\eea
Here, $F$ is the fraction of galaxies with redshift higher than the redshift of the cluster (according to the observed redshift distribution), while $\bar n$ is the mean galaxy density over the whole survey.  
Hence, averaging over the whole survey,
\be
\avg{\nbg} = F \bar n;\quad \avg{\nfg} = (1-F) \bar n.
\ee  
Further, $\efg$, $\ebg$, and $\ecl$ represent the foreground, background, and cluster ellipticity fields.  
The first two moments of $\ebg$ satisfy equations \ref{eq:mean_ellip} and \ref{eq:var_ellip}, while for the foreground
and cluster ellipticity fields we set
\bea
\avg{e_i} & = & 0 \\
\avg{e_ie_j} & = & \frac{1}{2}\sigma_e^2 \delta_{ij}.
\eea
In practice, cluster member galaxies are known to be radially aligned
(i.e., $\avg{\ecl_i}\neq 0$), but 
the extent of the radial ellipticities is less well known.
We defer this additional source of error to future work.
Upon plugging in, we find
\be
\avg{\DS} = \rbg \bar \Sc \mu^{q/2} g,
\ee
implying that $\DS$ is sensitive to only one of the three systematic parameters considered here.  The relative systematic error due to an uncertainty $\delta\rbg$ in $\rbg$
is simply
\be
\sigma_{\rm sys}(\ln \Delta\Sigma) = \frac{\delta \rbg}{\rbg} = \delta \rbg
\label{eq:sys_ds}
\ee
assuming our fiducial value $\rbg=1$.


\begin{figure}[t]
\epsscale{1.2}
\plotone{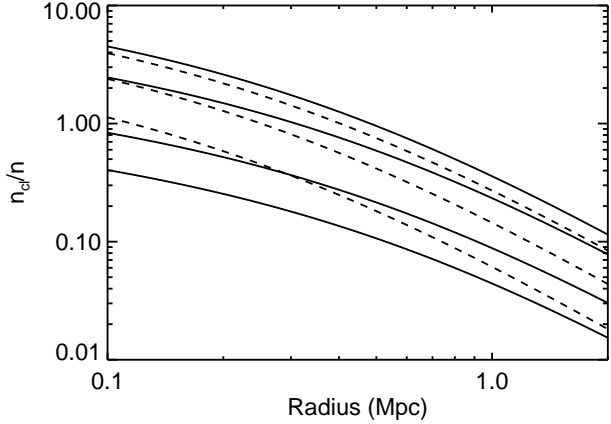}
\caption{{\it Top Panel:} The ratio $\bar \ncl/\bar n$ of the projected density of cluster galaxies to the mean galaxy density of the survey for 
a halo of mass $M=10^{15}\ \msun$ for a variety of redshifts.  From bottom to top, the redshift of the solid curves are $z=$0.1, 0.15, 0.3, and 
0.6.  From top to bottom, the redshift of the dashed lines are $z=$0.9, 1.2, and 1.5.  The redshift slice $0.2\lesssim z \lesssim 1.0$,
the ratio $\bar \ncl/\bar n$ can be larger than unity in the inner few hundred $\kpc$ for massive halos.
}
\label{fig:ratio}
\end{figure} 


Let us now turn our attention to $\DS'$, which takes the form
\be
\DS' = \bar \Sc \frac{x}{y}
\ee
where
\bea
x & = &  \sum_i \Delta\Omega \left[ \rbg \nbg \ebg_i + \rfg \nfg \efg_i  +  
\rcl \ncl \ecl_i  \right] \\
y & = &  \sum_i \Delta\Omega \left[ \rbg \nbg + \rfg \nfg + \rcl \ncl \right].
\eea
We will again ignore correction terms to the mean that go as $1/(\bar n A)$,
and set $\avg{\DS'}=\avg{x}/\avg{y}$.  We then arrive at
\bea
\avg{\DS'} & = &\bar \Sc g  \left [ 1 + \frac{\rfg}{\rbg} \frac{1-F}{\mu^{q/2}F} + \frac{\rcl}{\rbg} \frac{\avg{\ncl}}{\bar n} \frac{1}{\mu^{q/2}F} \right]^{-1} 
\eea
Here, $\avg{\ncl}$ is the mean number density of cluster member galaxies in the
stack considered.  
We find that $\avg{\DS'}$ is sensitive to all three of our systematics parameters $\rfg$, $\rbg$, and $\rcl$.  While these parameters
only appear as two independent ratios --- $\rfg/\rbg$ and $\rcl/\rbg$ --- in practice we still need to estimate all three parameters
independently.

But just how sensitive is $\DS'$ to systematic uncertainties?  As for $\DS$, we compute the relative systematic error due to a
small uncertainty in our systematics parameters.  Linearizing, we find
\be
\sigma_{\rm sys}(\ln \Delta\Sigma') =  - \delta \rfg \frac{1-F}{\mu^{q/2}F} - \delta \rcl \frac{\avg{\ncl}}{\bar n} \frac{1}{\mu^{q/2}F}.
\ee
Comparing the above expression to equation \ref{eq:sys_ds}, it is clear that $\DS'$ is more robust to systematics if the coefficient
of $\delta\rfg$ and $\delta\rcl$ are smaller than unity.  Inspecting the above expression we notice two important things:
first, the coefficients scale as $1/F$, so at large redshifts, $\DS$ is guaranteed to become more robust than $\DS'$.  Second,
the coefficient for $\delta\rcl$ scales as $\ncl/\bar n$ (dilution effect), 
which we expect to be larger than unity in the cores of clusters, but 
insignificant in the outskirts, so it is possible for $\DS$ to be more robust than $\DS'$ in the cores of clusters, but the opposite
be true in the outskirts. 

Figure \ref{fig:ratio} shows the ratio $\bar\ncl/\bar n$ for a halo of mass $10^{15}\ \msun$ as a function of radius
for a variety of redshifts.  The ratio has been estimated using the model detailed in Appendix \ref{app:cl_gals}.  As we
expected, this ratio is larger than unity in the inner few hundred $\kpc$ for $0.2 \lesssim z \lesssim 1.2$.  Note,
however, that this is true for very massive halos.  If we assume instead $M=10^{14}$, this ratio barely reaches
unity at $R=100\ \kpc$.  Note that the simple model for cluster members used 
here should only be taken as an order-of-magnitude estimate.  


\begin{figure}[t]
\epsscale{1.2}
\plotone{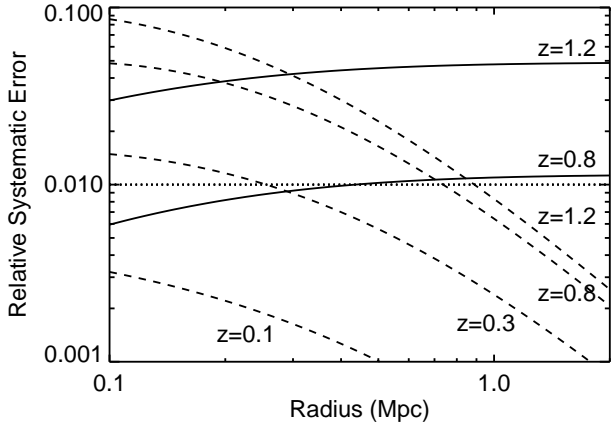}
\caption{Relative systematic error for both $\DS$ (dotted) and $\DS'$ 
(solid and dashed) as a function of radius for a halo of mass 
$M=10^{15}\ \msun$.  The solid curves show the error on $\DS'$ due to a $1\%$ 
systematic uncertainty $\delta r_{\rm fg}$ on the fraction of foreground 
galaxies included in the source galaxy population (the curves for $z=0.1, 0.3$ 
are below the minimum error plotted).  Dashed curves are the relative 
systematic error on $\DS'$ due to a $1\%$ error on $\rcl$, the fraction of 
cluster galaxies mistakenly included in the source population.  Finally, the 
dotted curves are the (redshift-independent) relative error on $\DS$ due to a $1\%$ systematic error 
$\delta r_{\rm bg}$ in the estimated fraction of background galaxies in the 
source population.}
\label{fig:sys}
\end{figure} 


Figure \ref{fig:sys} shows the relative systematic error for both the $\DS$ and $\DS'$ estimators assuming a systematic uncertainty
$\delta \rbg = \delta \rfg = \delta \rcl = 1\%$ (one can easily scale the
values in Figure~\ref{fig:sys} to other valus of $\delta r$).  
For the estimator $\DS$, the corresponding systematic error is constant, and mass and redshift independent.
The same is not true of $\DS'$.  We find that for massive halos, the error in $\DS'$ is typically dominated by dilution of the lensing signal by
cluster member galaxies.  This error becomes larger than the expected statistical error in $\DS$ at $z\approx 0.3$, though it affects only the
innermost few hundred $\kpc$ scales.  Moreover, this error is explicitly mass dependent, with less massive halos benefiting from smaller
systematic uncertainties.  In all cases, however, for $z\gtrsim 0.8$, the $1/F$ dependence of the systematic error on $\DS'$ makes this estimator
less robust than $\DS$.

From the above discussion alone, it is unclear if either of the two estimators we have considered here is superior to the other.  At very high masses and
high redshift, $\DS$ is likely to be preferable to $\DS'$, but this may be reversed as one moves towards lower redshifts and/or masses.  
Furthermore, choosing between $\DS$ and $\DS'$ also amounts to choosing 
between a spatially constant, mass-independent but likely redshift-dependent 
systematic in the case of $\DS$, and a significantly mass- and 
radius-dependent systematic in case of $\DS'$.  The cluster member dilution 
affecting $\DS'$ is frequently corrected for by multiplying the shear
profile measured with $\DS'$ by the observed source galaxy-cluster correlation
function \citep[e.g., ``boost factor'' in][]{mandelbaumetal05b}.  This is essentially
equivalent to using the estimator $\DS$ from the start.  
Finally, the answer as to which estimator is preferable
will also depend on which of 
the systematic uncertainties discussed here --- $\rbg$, $\rfg$, and $\rcl$  --- can be best controlled.


\subsection{Other Systematics}

In addition to the systematics discussed above, there are additional sources of systematic
uncertainty that can impact our results.  A few examples of such sources of systematic uncertainty are:
\begin{itemize}
\item Shear Systematics:  if shear is mis-estimated, this systematic will of course be carried over to the estimated cluster mass.
\item Miscentering: if clusters are miscentered, this can have a dramatic impact on the expectation value of the weak lensing signal in the cores of 
clusters \citep{johnstonetal07}.  This systematic will almost certainly dominate the uncertainty with which the concentration of galaxy clusters 
can be measured within cluster stacks, but its impact on cluster masses can be significantly reduced through careful analysis \citep{mandelbaumetal10}.
\item Source obscuration by foreground galaxies: occasionally, background galaxies will be perfectly aligned with foreground galaxies, and therefore
the latter
cannot be included in the weak lensing shear signal estimation.  This reduces the effective area of the annuli used to estimate shear, which in turn
impacts the expected number of galaxies in the annulus, thereby impacting globally normalized estimators.
\item Photometric redshift errors: the strength of the shear signal depends on the redshift of the source galaxy under consideration, which must
be estimated based on photometric data.  Consequently, scatter and catastrophic errors in photometric redshift estimates may have a significant 
impact on shear mass calibration experiments \citep[see e.g.][]{mandelbaumetal08}.
\end{itemize}

In light of this discussion, we reiterate that determining whether $\DS$ is superior to $\DS'$ or vice-versa will require empirical investigation,
and that due to these systematics, the forecasted precison for a
DES-like survey is best interpreted as the lower-limit for what will actually be realized.
That said, until one estimator may be conclusively shown to be superior to the other, 
estimating cluster masses using both estimators
should allow one to estimate the level of systematic uncertainty introduced by source galaxy selection,
which should be of tremendous utility in upcoming photometric surveys.


\section{The Impact of Stacked Weak Lensing on Cluster Abundance Experiments}

\subsection{Fisher Matrix and The Fiducial Model}

We now consider whether the statistical precision of the mean weak
lensing masses recovered from stacked weak lensing is sufficient to
significantly improve cosmological constraints in a DES-like survey
relative to the self-calibration expectation with Planck priors.  We address this
question by once again relying on the Fisher matrix formalism.
Specifically, we set the Fisher matrix for our experiment to the sum
of the standard self-calibration result plus an additional
contribution due to the stacked weak lensing data,
\be
F_{total} = F_{Planck}+F_{SC} + F_{WL}.
\ee

The Planck priors Fisher matrix is that provided by Hu and Ma (private communication).
For the self-calibration fisher matrix, we use the formalism described in detail in \citet{wuetal08} and \citet{wuetal10}.
Briefly, we assume that the survey area $5,000\ \deg^2$ is divided into $500$ patches of $10\ \deg^2$ each.
The observables in our experiment are the cluster counts in each of these $500$ patches for each mass and redshift bin.  
We adopt the same binning as for the weak lensing analysis (five bins per decade in mass, and redshift
slices of width $\pm\Delta z =0.05$).  

The model parameters we consider can be split into two categories, cosmological parameters,
and the nuisance parameters describing the observable--mass relation.  The cosmological parameters
and their fiducial values are given in the introduction, except that we add as free parameters $w_0$
and $w_a$.  These describe the equation of state of the dark energy $w=w_0+w_a (1-a)$ as a function of the cosmic expansion
factor $a$.  We will be primarily interested in how the Dark Energy Task Force \citep{detf06} figure of merit --- defined as the
the product of the eigenvalues of the $w$ and $w_a$ Fisher matrix --- changes
upon inclusion of the weak lensing data. 

In order to describe the observable--mass relation,
each cluster is assumed to have an observed ``mass'' $M_{obs}$
which represents its richness measurement.  We assume the observable--mass relation 
$P(M_{obs}|M_{true},z)$ is log-normal, with the mean and variance of
$\ln M_{obs}$ assumed to scale linearly with $\ln M$ and $\ln (1+z)$.
We write
\bea
\avg{\ln M_{obs}} & = & \ln M_0 + \alpha_M \ln \left( \frac{M}{M_{pivot}} \right) + \alpha_z \ln(1+z) \\
\sigma_{obs}^2 & = & \sigma_0^2 + \beta_M \ln  \left( \frac{M}{M_{pivot}} \right) + \beta_z \ln(1+z).
\eea
We set $M_{pivot}=7\times 10^{13}\ \msun$.\footnote{Note $\ln M_0$ and $\ln M_{pivot}$ are degenerate,
so we can fix one of them arbitrarily without loss of generality.}  The remaining set of parameters are allowed to vary,
and all of our results are marginalized over these observable--mass parameters.
We set the fiducial
value of our free parameters to those of an unbiased estimator in
$\ln M_{obs}$, so that $\ln M_0= \ln M$ and
$\alpha_M=0$ and $\alpha_z=0$.  Further, for our fiducial model we assume no evolution of the scatter with mass or redshift,
so that $\beta_M=\beta_z=0$.   We consider two possible values of the amplitude of the scatter $\sigma_0=0.2$ and $\sigma_0=0.5$,
corresponding to $20\%$ and $50\%$ scatter in $M_{obs}$ at fixed $M_{true}$.  

Our stacked weak lensing analysis allows us to introduce an additional set of observables in our analysis, namely the
mean mass $M_{WL}$ in bins of $M_{obs}$.   Since the mean mass estimates of the different bins are independent,
the weak lensing Fisher matrix is simply
\be
F_{WL} = \sum_{\mbox{all bins}} \frac{1}{\sigma_{WL}^2} \frac{\partial \avg{\ln M_{WL}}}{\partial p_i}  \frac{\partial \avg{\ln M_{WL}}}{\partial p_j}
\ee
where $\sigma_{WL}$ is the error in the mass estimated in section \ref{sec:staterr}.  To compute $\avg{\ln M_{WL}}$, we 
assume that weak lensing masses are unbiased, so that
\bea
\avg{M_{WL}|M_{obs},z} & = & \avg{M_{true}|M_{obs},z} \\
	& =  & \frac{1}{\bar N}\int dM_{true} dz\ M_{true} \frac{dn}{dM_{true}} \frac{dV}{dz} \avg{\phi}
\eea
where $\phi$ is the binning function in observed mass and redshift, $\avg{\phi}$ is the effective binning function
as a function of true halo mass,
\be
\avg{\phi|M_{true},z} = \int dM_{obs}\ \phi(M_{obs},z)P(M_{obs}|M_{true},z),
\ee
and $\bar N$ is the expected number of clusters,
\be
\bar N = \int dM_{true} dz\  \frac{dn}{dM_{true}} \frac{dV}{dz} \avg{\phi}.
\ee
A more detailed discussion of how to add this additional information to cluster forecasts is presented in \citet{wuetal10}.

Before moving on, we point out that since the error estimates $\sigma_{WL}$ of the weak lensing masses from section \ref{sec:staterr} assumed
clusters were binned according to their true masses, there is a small scatter-dependent correction to the predicted uncertainties.  Given that
we have not taken into account finite mass bin width effects (see section~\ref{sec:Fisher}),
we ignore these corrections in this section as well, and simply remind the reader that the corrections are expected to be small
since shape noise is larger than the intrinsic scatter in the mass.

\subsection{Results}


\begin{figure}[t]
\epsscale{1.2}
\plotone{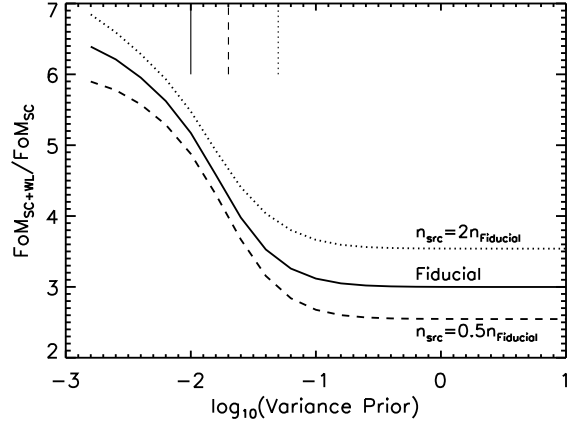}
\caption{The figure of merit for a DES-like cluster abundance experiment with stacked weak lensing mass calibration relative
to its self-calibration expectation, as function of the external prior $\Delta\sigma_0^2$
on the scatter in the mass-observable relation.  Planck priors are assumed.  The solid lines employ the forecasted errors arrived at in section \ref{sec:staterr}.
The dashed and dotted lines have a source density that is a factor of two lower (dashed) and higher (dotted) than that of our fiducial model.
The long tick marks
along the top axis correspond to a $5\%$ prior on the {\it scatter} assuming $\sigma_0=0.1$ (solid), $\sigma_0=0.2$ (dashed), 
and $\sigma_0=0.5$ (dotted).  The relative improvement in the figure of merit is nearly independent of the adopted value of $\sigma_0$,
particularly for broad priors.}
\label{fig:fom}
\end{figure} 


Figure \ref{fig:fom} shows the figure of merit of our DES-like survey including stacked weak lensing
mass calibration relative to the figure of merit obtained without this additional source of data.  We have
allowed for the possibility of a prior on the scatter parameters with
$\Delta \sigma_0^2 = \Delta \beta_M =\Delta \beta_z$.
The solid line assumes the weak lensing uncertainty for each mass and redshift bin $\sigma_{WL}$
estimated in section \ref{sec:staterr}, whereas the dashed and dotted assumed a source density
that is half and twice that of our fiducial model respectively.
Assuming no scatter prior, we find that the improvement in the Dark Energy Task Force figure of
merit is in the range $2.5-3.5$.    This is true for both the $\sigma_0=0.2$ and the $\sigma_0=0.5$
models: the relative improvement in the figure of merit is only weakly dependent on the assumed scatter,
which is why we only show one set of curves in Figure \ref{fig:fom} (those for $\sigma_0=0.5$).  
The fiducial value for the figure
of merit is, of course, different.  We have $FoM_{SC}=19$ for $\sigma_0=0.2$, while $FoM_{SC}=15$ for
$\sigma_0=0.5$.  

Additional priors on the scatter of the mass--observable relation may further increase the efficacy of stacked weak
lensing mass calibration, leading to improvements in the figure of merit as large as a factor of 8.  To do so, however
the priors need to be very tight.  Note that we have expressed
these priors in terms of the {\it variance} rather than the standard deviation.  For reference, the long
thin vertical tick marks a long the top x-axis in Figure \ref{fig:fom} correspond to a $5\%$ prior (i.e. $\Delta\sigma_0=0.05$ 
on the scatter, assuming $\sigma_0=0.1$ (solid), $\sigma_0=0.2$ (dashed), and $\sigma_0=0.5$ (dotted).  Note
that while strictly speaking we are only showing the relative improvement for the $\sigma_0=0.5$ case, the
curves for $\sigma_=0.1$ and $\sigma_0=0.2$ closely track that for $\sigma_0=0.5$.
It is evident from the Figure that accurate knowledge of the scatter 
can lead to significant further improvement in the figure of merit of the experiment.  

We have also considered how biases in the weak lensing masses could impact cosmological parameter estimation
in order to assess the level at which systematics need to be controlled.  Not surprisingly, we find that $\approx 2\%$
biases in mass --- which correspond to $\approx 1\sigma$ --- result in $\approx 1\sigma$ biases in the inferred
cosmological parameters.  Whether the recovered weak lensing masses can indeed be expected to be unbiased
at the $2\%$ level in real data, however, remains to be seen.


\section{Summary and Conclusions}

Weak lensing shear profiles and weak lensing peak finding tend to utilize the shear signal in different ways,
with shear profile mass calibration relying on locally normalized estimators ($\DS'$), and peak finding relying on globally normalized
estimators ($\DS$).  We have used both of these estimators to predict the precision to which the mean mass of galaxy cluster stacks
in a DES-like survey can be measured.  We find that for moderate redshift clusters ($z\lesssim 0.6$), the typical precision achieved
is $\approx 2\%$, with the two types of estimators having nearly identical statistical uncertainties.  A companion paper, \cite{paperB}, investigates 
similar issues in the case of weak lensing peak finding.

We also considered three sources of systematic biases for these measurements.  
The first is lensing bias, which we find affects $\DS$ but not $\DS'$.  For the former,
we find that including lensing bias corrections to the expectation value of $\DS$ is necessary to avoid significant
systematic biases in both halo mass and concentration.
We emphasize however that lensing bias corrections can be easily incorporated into the data analysis,
so this is not a particularly worrisome systematic.  Our results simply indicate that it is necessary to incorporate these corrections.

The second source of systematic uncertainty we considered are finite bin-width corrections, which affect both $\DS$ and $\DS'$.
Again, these corrections can in principle be
explicitly included when analyzing data, and our investigation only addresses whether doing so is necessary for practical purposes.  We
find that for logarithmic bins $\pm\Delta\log_{10} R \lesssim 0.04$, the 
systematic bias in mass from finite bin-width corrections is less than $1\%$. 
If one uses 5 bins per decade in radius ($\pm \Delta\log_{10} R = 0.1$), biases as large as $5\%$ in mass are possible.

The final systematic we consider here is fluctuations in the number
density of galaxies due to intrinsic clustering, which can affect the 
estimators through imperfect photometric redshift selections.  
Remarkably, we find that $\DS$ and $\DS'$ are
affected by this systematic in very different ways: $\DS$ is affected by the fraction of background galaxies missed by photometric redshift selection,
whereas $\DS'$ is affected by the fraction of foreground or cluster member galaxies that are included in the source population.   Whether $\DS$
or $\DS'$ is more robust depends on which of these systematics can be better controlled, a question which can only be empirically resolved.  In either case,
both estimators can be used to cross-check each other for the effects discussed
here, due to their significantly different systematics.  

Finally, having estimated the precision with which the mean cluster
mass of clusters stacks can be measured with a DES-like survey, we have investigated
how this measurement would impact the cosmological parameter constraints of a DES-like cluster abundance experiment.  We find that with our
fiducial assumptions, the figure of merit of such an experiment improves by 
a factor of 2.5--3.5, with larger increases possible if priors
on the scatter of the observable--mass relation can be derived from additional observations.  Furthermore, the improvement in the figure of merit induced by weak lensing
mass calibration is almost independent of
the magnitude of this scatter, and should thus
apply to a wide range of mass proxies.

\acknowledgements ER would like to thank Mike Jarvis, Matthew Becker, J\"org Dietrich, Scott Dodelson, Henk Hoekstra, Anja von der Linden, and Raul Jimenez,
for helpful discussions.  The authors would also like to thank Josh Frieman and J\"org Dietrich for comments on an earlier version of this manuscript
which helped to significantly improve the presentation.
ER is funded by NASA through the Einstein Fellowship Program, grant PF9-00068. 
HW is supported by the Gabilan Stanford Graduate Fellowship and the SLAC National Accelerator Laboratory.  
FS is supported by the Gordon and Betty Moore Foundation
at Caltech.

\bibliographystyle{apj}
\bibliography{mybib}

\newcommand\AAA[3]{{A\& A} {\bf #1}, #2 (#3)}
\newcommand\PhysRep[3]{{Physics Reports} {\bf #1}, #2 (#3)}
\newcommand\ApJ[3]{ {ApJ} {\bf #1}, #2 (#3) }
\newcommand\PhysRevD[3]{ {Phys. Rev. D} {\bf #1}, #2 (#3) }
\newcommand\PhysRevLet[3]{ {Physics Review Letters} {\bf #1}, #2 (#3) }
\newcommand\MNRAS[3]{{MNRAS} {\bf #1}, #2 (#3)}
\newcommand\PhysLet[3]{{Physics Letters} {\bf B#1}, #2 (#3)}
\newcommand\AJ[3]{ {AJ} {\bf #1}, #2 (#3) }
\newcommand\aph{astro-ph/}
\newcommand\AREVAA[3]{{Ann. Rev. A.\& A.} {\bf #1}, #2 (#3)}

\appendix


\section{Source Clustering}
\label{app:src_clustering}

In the main section of this paper, we assumed source galaxies are randomly distributed in the sky.  We now investigate whether source clustering can have
an impact on our results.  Source clustering implies that equation \ref{eq:poisson} must be replaced by 
\be
\avg{\delta_i\delta_j} = \delta_{ij} \frac{1}{\mu^{q/2}\bar n \Delta \Omega} + \xi_{ij}
\ee
where $\xi_{ij}$ is the projected galaxy--galaxy correlation function.  Note that in general this term will correlate different radial bins,
so that if $\DS_\alpha$ and $\DS_\beta$ are the estimators at radial bins $\alpha$ and $\beta$, the new term in the covariance 
matrix is given by
\be
C_{\alpha\beta} = \bar \Sc^2 \mu_\alpha^{q/2}\mu_\beta^{q/2} g_\alpha g_\beta V_{\alpha\beta}
\label{eq:cab}
\ee
where
\be
V_{\alpha\beta} = \frac{1}{A_\alpha A_\beta} \sum_{ij} \Delta\Omega^2 \xi_{ij}W^\alpha_i W^\beta_j,
\ee
and $A_\alpha$ and $A_\beta$ are the area of the annulus $\alpha$ and $\beta$ respectively.
Note we have used the thin annulus approximation to set $\mu$ and $g$ constant within each annulus.  
Taking now the continuum limit, we arrive at
\be
V_{\alpha\beta} = \frac{1}{A_\alpha A_\beta} \int \frac{d^2l}{(2\pi)^2}\ P(l)|W_\alpha^*(l) W_\beta(l)|.
\ee

Before going any further, it is worth taking a second to compare equation \ref{eq:cab} with equation \ref{eq:dserr}.
Specifically, note that shape noise is explicitly dependent on the source density, whereas source clustering is not.
This implies that at a sufficiently high source density, source clustering must dominate.  On the other hand,
source clustering scales as $g^2$, whereas shape noise scales as $\sigma_e^2$.  More precisely, source clustering will be relevant if
$g_\alpha^2 V_{\alpha\alpha}$ is comparable to $\sigma_e^2/2\bar n A$.  Thus,
unless the source density is quite large, we expect shape noise to dominate.  
What follows is a quantitative confirmation of this expectation.

To do so, we need to begin by estimating $V_{\alpha\beta}$, which in turn requires that we compute the projected source galaxy power spectrum.
Let then $n_{co}(\bx)$ be the comoving galaxy density field.  Assuming flatness, the corresponding projected galaxy density field is 
\be
n(\vec \theta) = \int dz\ n_{co} \chi^2 \frac{d\chi}{dz} H(z-z_L)
\ee
where $H$ is a step function that selects only galaxies at redshift larger than the lens redshift of interest $z_L$.
The fluctuations in the source density field are therefore given by
\be
\delta(\vec \theta) = \int d\chi\ g(\chi) \delta_g(\bx)
\ee
where $\delta_g(\bx)$ is the 3D galaxy density contrast, and 
\be
g(\chi) = \frac{1}{\int dz\ f(z)H(z-z_L)} f(z) \left(\frac{d\chi}{dz}\right)^{-1} H(z-z_L).
\ee

Using Limber's approximation, the angular power spectrum of the source density field is related to the three dimensional galaxy power spectrum
via
\be
P(l) = \int d\chi\ \frac{g^2(\chi)}{\chi^2} P_{3D}(l/\chi,z).
\ee
All that remains is to specify the 3D galaxy power spectrum.  To do so, we assume a constant scale-independent bias $b=1$ relative to the non-linear
matter power spectrum $P_{mm}(\bm{k},z)$, which we compute using \citet{smithetal03} as implemented in the CAMB package \citep{lewisetal99}.

Having determined the projected source power spectrum, we need to compute the mean power over the filter functions $W_\alpha$.  We assume
$W_\alpha$ is a top hat in radius going from $\log_{10} R \in [\log_{10} R_c-\Delta,\log_{10} R_c+\Delta]$, with $\Delta = 0.02$ as per
our fiducial assumptions.  The Fourier transform of $W_\alpha$ is therefore
\bea
W_\alpha(l) & = & \int d^2\theta\ W_\alpha(\theta) \exp(i\bm{l} \cdot \bm{\theta}) = 2\pi \int_{\theta_{min}}^{\theta_{max}} d\theta\ \theta J_0(l\theta) \\
	& = & \frac{2\pi}{l^2} \left[ u_{max}J_1(u_{max}) - u_{min}J_1(u_{min}) \right]
\eea
where $u= Rl/d_A$ and $d_A$ is the angular diameter distance to the redshift of the lens.  We now have all the ingredients necessary for computing
the source clustering error.


\begin{figure}[t]
\epsscale{0.75}
\plotone{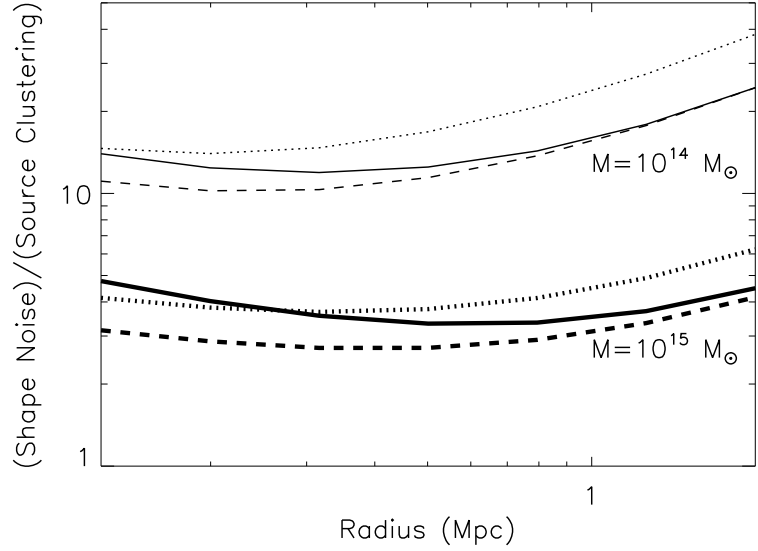}
\caption{Ratio of the shape noise error considered in the main body of the manuscript to our estimated source clustering error using our fiducial
assumptions for the estimator $\DS$.  The ratio is shown for halos of mass $M=10^{14}\ \msun$ (thin lines) and $M=10^{15}\ \msun$.  Different
lines correspond to different redshifts, namely $z=0.1$ (solid), $z=0.5$ (dashed), and $z=0.9$ (dotted).  In all cases, we assume a concentration
$c=5$.  We conclude that for the source densities expected for the DES source clustering is a subdominant source of noise.
}
\label{fig:error_ratio}
\end{figure} 


Figure \ref{fig:error_ratio} compares the ratio between the shape noise term used in the main section of the paper, and the diagonal contribution
$C_{\alpha\alpha}$ to the covariance matrix due to source clustering.   Note we plot the ratio of the errors (i.e. the square root of the variance) rather
than the ratio of the variance.  The ratio of the noise terms is estimated for halos of mass $M=10^{14}\ \msun$ (thin lines) and $M=10^{15}\ \msun$ (thick lines)
for three different redshifts, $z=0.1$ (solid), $z=0.5$ (dashed), and $z=0.9$ (dotted).  We can see that source clustering is always significantly
smaller than the corresponding shape noise terms, so that it can be safely neglected in our forecast.    It is worth remarking, however,
that the ratio of these two noise terms is dependent on the assumed radial bin width, with broader bins leading to smaller ratios reflecting
the decreased shape noise.  Figure \ref{fig:error_ratio} assumes our fiducial bin width $\pm \Delta \log_{10}=0.02$.  If we were to use relative broad radial bins
of width $\pm \Delta \log_{10} R = 0.1$, the error ratio can drop to a factor of 3 for the $10^{15}\ \msun$ halos and down to a factor of $\sim 10$
for the $10^{14}\ \msun$ halos.   Thus, even when employing broad radial bins, the errors are dominated by shape noise rather than by source clustering.


\section{The Mean and Variance of $\DS'$}
\label{app:dsprime}

The estimator $\DS'$ takes the form $\DS'=\bar \Sc x/y$ where
\bea
x & = & \sum_i \Delta\Omega (1+\delta_i)e_i W_i \\
y & = & \sum _i \Delta\Omega (1+\delta_i) W_i.
\eea
We write $x=\bar x +\Delta x$ and $y=\bar y +\Delta y$, and use the binomial expansion to 
solve for $\DS'$ assuming $\Delta y/\bar y \ll 1$ to expand up to second order.  We find
\be
\DS' = \Sc  \frac{\bar x}{\bar y} \left[ 1 + \frac{\Delta x}{\bar x} - \frac{\Delta y}{\bar y} - \frac{ \Delta x \Delta y }{\bar x \bar y} + \frac{ \Delta y^2 }{\bar y^2} \right].
\label{eq:dsxy}
\ee
Upon taking the expectation value, and using the fact that $\bar x = Ag$ and $\bar y = A$, we arrive at
$\avg{\DS'}=\Sc g$, as per equation \ref{eq:avgdsprime}.   Squaring equation \ref{eq:dsxy}, we find
\be
(\DS')^2 = \Sc^2 \left( \frac{\bar x}{\bar y} \right)^2 \left[ 1 + \frac{\Delta x^2}{\bar x^2} + \frac{3\Delta y^2}{\bar y^2} - \frac{4\Delta x\Delta y}{\bar x\bar y} \right].
\ee
To compute the expectation value, we use the fact that $\bar x=gA$, $\bar y = A$, and
\bea
\avg{\Delta x^2} & = & \left( g^2+ \frac{1}{2}\sigma_e^2 \right) \frac{A}{\mu^{q/2} \bar n} \\
\avg{\Delta x \Delta y} & = & \frac{gA}{\mu^{q/2} \bar n} \\
\avg{\Delta y^2} & = & \frac{A}{\mu^{q/2} \bar n}.
\eea
The expectation value of $\avg{\DS'^2}$ simplifies to
\be
\avg{(\DS')^2} = \Sc^2 g^2 \left[ 1 + \frac{1}{\mu^{q/2} \bar n A} \frac{\sigma_e^2}{2g^2} \right],
\ee
which leads directly to equation \ref{eq:vardsprime}.


\section{A Model for Cluster Galaxies}
\label{app:cl_gals}

In order to compute this systematic error due to $\delta\rcl$ we must first estimate the
ratio $\bar \ncl/\bar n$.  For this, we derive an order-of-magnitude estimate 
as follows: let $\epsilon$ be the fraction of galaxies within a narrow
redshift slice $\pm \Delta z$ about the redshift of the cluster, and $\pm L$ be the corresponding {\it physical} width
of the slice.  These two quantities are related to the redshift width of the slice via
\be
\epsilon =  \frac{1}{\bar n} \frac{d\bar n}{dz} 2\Delta z  = \frac{f(z)}{\int_0^\infty dz'\ f(z')} 2\Delta z
\ee
where $f(z)$ is given by equation \ref{eq:fz}, and
\be
L = \frac{\Delta z}{1+z} c\Hinv.
\ee
Now, the mean three dimensional galaxy density $\rho_g$ within the redshift slice is
\be
\bar \rho_g = \frac{\epsilon \bar n}{2LD_A^2}.
\ee
Letting $\delta_g$ be the galaxy density contrast field, the projected cluster galaxy density is given by 
\be
\bar \ncl = D_A^2 \int \frac{d\chi}{1+z}\ \rho_g = D_A^2 \bar \rho_g \int \frac{d\chi}{1+z} (1+\delta_g) = \frac{\epsilon \bar n}{2L} \int  \frac{d\chi}{1+z} (1+\delta_g)  .
\ee
Assuming galaxies trace mass, we can set $\delta_m=\delta_g$,  and therefore
\be
(1+\delta_g) = (1+\delta_m) = \frac{\rho_m}{\bar \rho_m(z)}.
\ee
Replacing in our expression for $\bar \ncl$,
\be
\bar \ncl =  \frac{\epsilon \bar n}{2L} \int \frac{d\chi}{1+z} \frac{\rho_m}{\bar \rho_m(z_L)} =  \frac{\epsilon \bar n}{2L\bar \rho_m(z)} \Sigma
\ee
Inserting our expressions for $\epsilon$ and $L$, and setting $\bar \rho_m = \Omega_m \rho_c (1+z)^3$ we arrive finally at
\be
\frac{\bar \ncl}{\bar n} = \frac{f(z)}{\int_0^\infty dz'\ f(z')} \frac{\Sigma}{(1+z)^2\Omega_m\rho_c c\Hinv(z)}.
\ee

\end{document}